\documentclass[usenatbib]{mn2e}

\usepackage{verbatim}
\usepackage{graphicx,epsfig}

\title[Host galaxies of long gamma-ray bursts
]{Host galaxies of long gamma-ray bursts
in the {\it Millennium Simulation}}
\author[N. E. Chisari, P. B. Tissera, L. J. Pellizza]{N. E. Chisari$^{1}$\thanks{E-mail:
nchisari@princeton.edu (NECh); current address: Department of Astrophysical Sciences, Princeton University, Princeton, NJ 08544.}, P. B. Tissera$^{1,2}$ and L. J. Pellizza$^{1,2}$\\
$^{1}$ Instituto de Astronom\'{\i}a y F\'{\i}sica del Espacio, Casilla de Correo 67, Suc. 28, 
1428, Buenos Aires, Argentina\\
$^{2}$ Consejo Nacional de Investigaciones Cient\'{\i}ficas y T\'ecnicas, CONICET, Argentina}

\begin{document}

\date{submitted 2009 November 24, accepted 2010 June 09}

\pagerange{\pageref{firstpage}--\pageref{lastpage}} \pubyear{2009}

\maketitle

\label{firstpage}

\begin{abstract}
In this work, we investigate the nature of the  host galaxies  of long
Gamma-Ray bursts (LGRBs) using a galaxy catalogue constructed from the {\it Millennium Simulation}. 
We developed a LGRB synthetic model based on the hypothesis that these events originate at the end of the life of
massive stars following the collapsar model, with the possibility of including a constraint on the metallicity of the progenitor star. 
A complete observability pipeline was designed to calculate a probability estimation for a galaxy to be observationally identified as a host
for LGRBs detected by present observational facililties. 
This new tool  allows us to  build an observable host galaxy catalogue which is required to reproduce
the current stellar mass distribution of observed hosts.
This observability pipeline predicts that the minimum mass for the
progenitor stars should be $ \sim 75 $ M$_\odot$ in order to be able to reproduce BATSE observations.
 Systems in our  observable catalogue are able to reproduce the observed properties of host galaxies, namely 
stellar masses, colours, luminosity, star formation activity and metallicities as a function of redshift. At $ z >2$,
 our model predicts that the observable host galaxies would be very similar to
the global galaxy population. We found that $\sim 88$ per cent of the observable host galaxies
with mean gas metallicity lower than $0.6\, {\rm Z_\odot}$ have
stellar masses in the range  $  10^{8.5}$--$10^{10.3}$M$_\odot$ in
excellent agreement with observations. Interestingly, in our model
observable host galaxies remain mainly within this mass range regardless of redshift,
since lower stellar mass systems would have a low probability of being observed while more massive ones
would be too metal-rich. Observable host galaxies are predicted to
preferentially inhabit dark matter haloes  in the range  $10^{11}$--$ 10^{11.5}$M$_\odot$, with a weak dependence on  redshift. They are also found to preferentially map different density environments
at different stages of evolution of the Universe. At high redshifts, the observable host galaxies 
are predicted to be located in similar environments as the global galaxy population but to have a slightly higher probability to have a close companion.

\end{abstract}

\begin{keywords}
gamma-rays: bursts -- methods: numerical -- stars: formation -- galaxies: evolution, interactions.
\end{keywords}

\section{Introduction}
\label{int}

Gamma-ray bursts (GRBs) are brief pulses of $\gamma$-ray radiation observed on
average once a day at random directions in the sky. They are the brightest
sources in this region of the electromagnetic spectrum, and they have been
systematically studied in the past two decades \citep[e.g.][and references
therein]{Mes06}. The origin of GRBs is cosmological, as determined from
the measurement of their redshifts \citep*[e.g.][]{Met97,Blo01,Blo03}, and in many cases, their host galaxies 
have been identified \citep*[see][and references therein]{LeF03,Sav09}. Their distances imply the
release of large amounts of energy ($\sim10^{51}\,{\rm ergs}$) in a short
time-scale, which suggests that the origin of these phenomena could be
associated to the accretion of matter onto a compact object \citep{Woo93,Fry99,McF99,Pan01,Fra01}.

Two populations of GRBs are apparent from the distribution of their duration
\citep{Kou93}. Those lasting less than $2\,{\rmn s}$ are known as short GRBs,
while longer events are called long GRBs (LGRBs). These are more frequently
observed and more precisely located, which makes their properties better known.
LGRBs are always found in host galaxies with ongoing star formation activity
\citep*{LeF03,Chr04,Pro04}, and some of them were observed to be
associated with Type Ib/c core-collapse supernovae \citep{Gal98,Hjo03}. These
observations suggest that the progenitors of LGRBs are massive stars. Stellar
evolution models have tried to provide a consistent scenario where a LGRB can
develop. The result is the so called {\em collapsar model} \citep*{Woo93,Fry99},
in which this phenomenon is produced during the collapse of a massive star,
due to the accretion of part of the envelope onto the recently formed black
hole. Nevertheless, the properties of LGRB progenitors (mass, metallicity,
rotation velocity, binarity, etc.) are still a matter of discussion. 
LGRBs have been associated to single massive stars \citep{Woo93}, perhaps in
metallicity biased environments \citep*{Hir05}, and also to stars in binary
systems \citep{Bis07}. 

The importance of understanding the nature of the LGRB progenitors is beyond the
interest  of  only  stellar evolution and black hole formation. Given their
connection to massive stars and their large luminosities, LGRBs might be a
powerful tool to investigate the star formation in the early Universe, at
redshifts for which standard tools become ineffective \citep[e.g.][]{Wij98}. The knowledge of the
properties of the stellar progenitors would allow the assessment of possible
biases originated when LGRBs are used as tracers of star formation. Several
authors suggest that the cosmic LGRB rate does not follow the star formation
rate \citep*{Dai06,Sal07}. The interpretation of their data requires the
assumption of a differential evolution of the comoving LGRB rate density with
respect to the comoving star formation rate (SFR) density. The same conclusion arises from the
work of \citet{Wol07}, who show that the luminosity function of LGRB hosts
differs from that of core-collapse supernovae hosts, which are considered
unbiased tracers of star formation. According to these authors, the LGRB host
luminosity function can be reproduced by requiring LGRB progenitors to have
metal abundances lower than that of the Sun.	

An indirect procedure to investigate LGRB progenitors is to characterise the
stellar populations of their host galaxies. Given the connection between LGRBs and star
formation, galaxy formation models can be used to assess the validity of LGRB
progenitor models, comparing their predictions about the properties of these
stellar populations to host galaxies observations. Recent works such as that of
\citet{Kau04}, suggest that there is a strong relation between star
formation and nuclear activity in a galaxy with its environment, at constant
stellar mass. Hence, the environment of host galaxies might constitute an independent
probe for progenitor models. 
Observational works on host galaxies environment are still
inconclusive on whether host galaxies inhabit regions with certain characteristics \citep{Fyn02,Jak05,Tho08}.
\citet{Bor04} suggest that host galaxies are field galaxies, while \citet*{Wai07} find
that a considerable fraction of their sample of 42 host galaxies shows evidence of
interaction with other galaxies. The question of whether there is a connection
between the occurrence of a LGRB and the local density of galaxies remains
unanswered.

Since galaxy formation is a highly non-linear process, the properties of host galaxies
predicted by different progenitor scenarios are better studied by means of
cosmological numerical simulations \citep{Kat91,Nav93,Mos01,Spr03,Sca05,Sca06}.
\citet*{Cou04} use hydrodynamical simulations of structure formation to identify
a galaxy population whose properties reproduce the observed ones. According to
these authors, the luminosity distribution of host galaxies is reproduced if galaxies are
required to have high star formation efficiency. However, their simulations do
not make any prediction on the properties of the LGRB progenitors.
\citet{Nuz07} developed a Monte Carlo code to simulate the production of LGRBs
in hydrodynamical simulations of galaxy formation, assuming that their
progenitors are massive stars as proposed by the collapsar model.
These authors follow the  LGRB production  as the structure forms and
evolves in the Universe.
Their
results suggest that LGRB progenitors would be low metallicity stars ($Z < 0.3
Z_\odot$), and hence LGRBs would be biased tracers of star formation principally at low
redshift. However, their simulations explore a small volume of $(10
{h^{-1}\,\rm Mpc})^3$  dominated by field galaxies.
Recently, \citet{Cam09}  have constructed a simulated population of host galaxies based on
the semi-analytic model of galaxy formation of \citet{DeL07} applied to
cosmological simulations developed by \citet{Wan08}. These simulations describe
a larger volume of the Universe (a square box of $125 {h^{-1}\,\rm Mpc}$ in a
side). \citet{Cam09} explore three LGRB progenitor models based
on the collapsar model, one of them assuming a given stellar mass value and an age cut-off and  the other
two also  including  metallicity cut-offs. These authors find that
models with very low
metallicity progenitors ($Z < 0.1 Z_\odot$) could explain the luminosities,
colours and metallicities of the observed host galaxies. Their results support previous
claims of LGRBs being biased tracers of the star formation.

Clearly, the use of larger simulations with better resolution and more detailed
descriptions of the physical processes driving the dynamical, chemical and
star formation histories of galaxies allows a better modelling of the
host galaxies of LGRBs.  However, an important aspect that should not be
disregarded is that, for a proper comparison with observed host galaxies, the effects of
the detectability of these galaxies must be taken into account. And
this is the main contribution of our work where we developed an
observational pipeline which allows us to mimic, at least in part,
biases affecting the observations. With this new tool we build an
observable host galaxy catalogue which can be more fairly compared to
current available observations. 

In this work, we develop a semi-analytical model for the  host galaxies
of LGRBs adopting the
collapsar model for the  progenitor stars. We apply it to one of the largest
cosmological simulations available, the {\em Millennium Simulation}
\citet{Spr05a}. The {\em Millennium Simulation} follows the evolution of dark
matter in a box of $500 {h^{-1}\,\rm Mpc}$ in a side, hence providing a
cosmologically representative volume a factor of 64 larger than that of
\citet{Wan08} and factor of $\sim 10^5$ larger than that of
\citet{Nuz07}. We take the galaxy catalogue built up by \citet{DeL07}, who
describe the evolution of baryonic matter including star formation, active
galactic nuclei and supernova feedback. Using the star formation and chemical
properties of galaxies in this catalogue to implement different LGRB progenitor
scenarios, we determine the properties of the overall host galaxy population and its
environment. We compute the detectability of the host galaxies in order to compare the
predictions of these scenarios to the observations compiled by \citet{Sav09}.

This work is organized as following. In Section~\ref{sim}, we briefly describe the
cosmological simulation and semi-analytic galaxy catalogue used in our work.
In Section~\ref{mod}, we develop our implementation of LGRB progenitor
scenarios. Sections~\ref{hgs} and \ref{env} show our results on host galaxy properties and
environment, respectively. Finally, Section~\ref{con} presents our conclusions.

\section[]{Millennium Simulation}
\label{sim}

The {\em Millennium Simulation} \citep{Spr05a} is one of the largest
cosmological simulations of structure formation publicly available. It
describes the evolution of the cold dark matter while the baryonic matter can be 
included via semi-analytic models on top of the numerical simulation. The
{\em Millennium Simulation} describes the formation of structure tracking 
$\sim10^{10}$ dark matter particles of mass $8.6 \times 10^8 h^{-1}\,{\rm M_{\odot}}$
distributed in a cubic region of 500 $h^{-1}\,{\rm Mpc}$ side, using the
$\Lambda{\rm CDM}$ cosmogony. The adopted cosmological parameters are
$\Omega_{\rm m}=0.25$, $\Omega_{\rm b}=0.045$, $\Omega_{\Lambda}=0.75$,
$\sigma_8=0.9$ and $n=1$, where the Hubble constant is $H_0=100{h\,\rm km}
\,{\rm s}^{-1}\,{\rm Mpc}^{-1}$ with $h=0.73$. These parameters are chosen in
consistency with results obtained by the joint analysis
of the Two-degree Field Galaxy Redshift Survey
(2dFGRS\footnote{http://www.mso.anu.edu.au/2dFGRS/}) and the Wilkinson
Microwave Anisotropy Probe (WMAP) data \citep{San06}.

The simulation was performed using a modified version of GADGET-2
hydrodynamical code \citep{Spr05b}. A {\em friends-of-friends} (FOF) algorithm
identified non-linear dark matter haloes within the simulation in an automatic
manner. Two particles belong to the same halo if their separation is less than
$0.2$ times the mean separation between particles. Only groups of more than
$20$ particles are identified as haloes. This restriction sets a lower limit on
the mass of a halo of $1.72 \times 10^{10} h^{-1}\,{\rm M_{\odot}}$. Substructures
orbiting within each virialized halo are identified applying a SUBFIND
algorithm \citep{Spr01}. Each dark matter halo has a central (Type 0) galaxy
and one or more satellite galaxies. Satellites were at some point central
galaxies of smaller haloes which suffered a merger with the halo their
currently inhabit. There are two types of satellite galaxies, Type 1 galaxies
are located at the centre of a subhalo associated with a FOF group, while Type
2 galaxies have lost their dark matter subhalo after falling  onto a more
massive halo.

Mean properties of synthetic galaxies (e.g. SFR, stellar mass,
gas mass, metallicity) are obtained by applying semi-analytic models to the
structure formation simulation \citep{Whi91,Col91,Lac91}. Essentially,
semi-analytical models describe the collapse of baryonic matter following dark
matter haloes. After collapsing, the gas cools and gravitational instabilities
produce episodes of star formation. To reproduce observations, models
incorporate photoionization processes in the intergalactic medium, the growth
of supermassive black holes during galactic mergers, supernova and AGN
feedback, star formation rate enhancement in mergers, and formation of heavy
elements for each formed stellar population. The galaxy catalogue we use in
this work is that of \citet{DeL07}.

There is recent evidence for an excessive reddening of Type 2
galaxies 
 \citep*{Wei06,Per09} which might be due to the
poor physical treatment of this type of galaxies. In order to avoid spurious
trends, we exclude Type 2 galaxies from our analysis in Section~\ref{mod} and
Section~\ref{hgs}. However, Type 2 galaxies ought to be included in the
environmental analysis performed in Section~\ref{env} in order to correctly trace
the underlying mass distribution.

Hence, the catalogue of \citet{DeL07} provides us with the spatial galaxy
distribution, their stellar masses, dark matter haloes, star formation
activity, colours, luminosities and mean metallicities. Regarding the latter,
the public catalogue makes available the mean metallicity of the cold gas
component and of the stellar population as a whole (i.e. averaged over new and
old stars in a galaxy). Therefore, as a proxy for the metallicity of the
LGRB progenitors, we take the mean metallicity of the cold gas component at the
time these stars were born. Because the \citet{DeL07} model assumes
instantaneous recycling, this metallicity is slightly higher than that of the
newly born stars, but this correction is negligible
compared to the uncertainties in the measured metallicities  \citep[e.g.][]{Cam09}.

\section[]{Scenarios for LGRB progenitors}
\label{mod}

\subsection{Intrinsic LGRB rate}

We consider two scenarios for LGRB production, both of them based on the
collapsar model \citep{Woo93,McF99,Fry99}. In our scenario I we take as LGRB
progenitors all stars above a certain minimum mass $m_{\rm min}$, with no
other restriction whatsoever. In this scenario LGRBs are unbiased tracers of
star formation. Accordingly, we obtain the intrinsic LGRB rate in a given
galaxy $g$ at a particular redshift $z$ as

\begin{equation}
\label{rateint}
r_{\rm GRB}(g,z)= {\rm SFR}(g,z) \frac {\int_{m_{\rm min}}^{100\,{\rm M_{\odot}}}
\xi(m)\,dm}{\int_{0.1\,{\rm M_{\odot}}}^{100\,{\rm M_{\odot}}} m \xi(m)\,dm},
\end{equation}

\noindent
where $0.1\,{\rm M_{\odot}}$ and $100\,{\rm M_{\odot}}$ are the lower and upper mass cut-offs of
the IMF $\xi(m)$ given by \citet{Cha03}. According to Eqn.~\ref{rateint},
the production of LGRBs is not delayed with respect to the starburst that
created the progenitor stars, which is justified because in the collapsar model
only massive stars are LGRB progenitors.

In our scenario II, we assume that only massive stars ($m > m_{\rm min}$) below
some metallicity threshold ($Z_{\rm C}$) are able to produce LGRBs, as in the
progenitor models of \citet{Hir05} or \citet*{Yoo06}.  As explained before, we
take the mean metallicity of cold gas in each galaxy as representative of the
metallicity of the progenitor stars. The consequences of this hypothesis will
be discussed in later sections. In this scenario, the intrinsic LGRB rate is
given by Eqn.~\ref{rateint} for galaxies $g$ with metallicity $Z_g <
Z_{\rm C}$, and is null for those with $Z_g \geq Z_{\rm C}$. Three
realizations of this scenario were computed, adopting $Z_{\rm C} = 0.1,\,
0.3,\, 0.6\, {\rm Z_{\odot}}$ (scenarios II.1, II.2 and II.3, respectively).

\subsection{Observed LGRB rate}
\label{obsrate}

Given a metallicity threshold, we adjusted the only free parameter of our
scenarios ($m_{\rm min}$) to reproduce the LGRB rate measured by the BATSE
experiment. For this purpose, we took the off-line GRB search of \citet{Ste01},
which detected 3475 GRBs with $T_{90} > 2\,{\rm s}$ during a live-time of
$6.37\,{\rm yr}$ (70 per cent of the $9.1\,{\rm yr}$ that BATSE was active),
scanning 67 per cent of the sky. This translates into a full-sky LGRB rate of
$814\,{\rm yr}^{-1}$ above the threshold of the off-line search. The choice of
this particular experiment was motivated by its good statistics and the
availability of an accurate model for its detection efficiency.

To obtain the observable rate predicted by a given scenario, we first compute
the comoving LGRB rate density for each redshift as

\begin{equation}
\label{e:comovrate}
n_{\rm GRB}(z) = V^{-1} \sum_g r_{\rm GRB}(g,z),
\end{equation}

\noindent
where $V = (500 \ h^{-1}\ {\rm Mpc})^3$ is the comoving volume of the {\it
Millennium Simulation}. The full-sky observed LGRB rate is then

\begin{equation}
\label{ratebatse}
R_{\rm GRB} = \int_0^{z_{\rm max}}\frac{n_{\rm GRB}(z)}{(1+z)} p_{\rm det,BATSE}(z)
\frac{dV}{dz},
\end{equation}

\noindent
where

\begin{equation}
\label{dvdz}
\frac{dV}{dz} = \frac{4 \pi c d_L^2(z)}{H(z)(1+z)^2}
\end{equation}

\noindent
is the derivative of the comoving volume with respect to $z$ at fixed solid
angle, $z_{\rm max}$ is the maximum redshift of the simulation, $d_L$ the
luminosity distance and $H$ the Hubble constant at redshift $z$, and
$p_{\rm det,BATSE}$ the probability of detecting a LGRB at redshift $z$ with BATSE (i.e. the
probability that a particular LGRB has a peak flux above the experiment
threshold). The integration was performed numerically, and the described
procedure iterated over  $m_{\rm min}$, until agreement with the observed rate
was attained.

The value of $p_{\rm det,BATSE}(z)$ depends on the LGRB luminosity function and
spectrum. It was computed using a Monte-Carlo scheme to simulate, at a given
redshift, a large population of LGRBs with different luminosities and spectra. 
Then, their photon fluxes in the BATSE energy band were estimated. For each
LGRB in this population, a second Monte-Carlo procedure rejected those events
which would be undetectable, taking into account the trigger efficiency of the
\citet{Ste01} off-line search. If the LGRB emission were isotropic, the fraction of retained LGRBs would directly give
the detection probability due to the off-line search threshold. To account for
the beamed emission of LGRBs \citep[e.g.][]{Yon05}, and assuming that the
distribution of jet opening angles is independent of the LGRB luminosity,
spectrum and redshift, we obtain $p_{\rm det,BATSE}$ by multiplying this fraction by
the mean beaming fraction of the jets $p_{\rm jet} = 10^{-3}$. The luminosity function
and spectral parameters distributions were taken from \citet{Dai06}.

This procedure not only allows us to mimic the observational process but also
has provided us with estimates of the minimum mass for the progenitor stars.
The values of $m_{\rm min}$ obtained, in solar masses, are $91.4 \pm 0.1$,
$13.9 \pm 0.2$, $44.3 \pm 0.4$, and $76.0 \pm 0.3$ for scenarios I, II.1, II.2
and II.3, respectively; the uncertainties reflect Poissonian errors in the
number of observed LGRBs (see also Table 1).

\begin{table}
  \begin{center}
  \caption{Main characteristics of our scenarios for the simulated
    host galaxies of LGRBs. Column (1) gives the name
    of the scenario.  Column (2) lists the minimum stellar mass obtained
    for progenitor stars of LGRBs to reproduce BATSE observations,
    adopting \citet{Cha03} IMF. Column (3)
    gives the cold gas maximum metallicity cut-off. Column (4) shows
    ratio between  the percentage of simulated host galaxies over the
    mass  range $\approx 10^{8.5-10.3} M_\odot$  and the corresponding value 
    obtained from the sample of \citet{Sav09}.}
\label{tab1}
  \begin{tabular}{|c|c|c|c|}\hline
 {Scenarios}  & { $m_{\rm min}$ (${\rm M_\odot}$)} & { $Z_{\rm C}$}  &  {R$_{\rm M}$}  \\\hline
 I   &  $91.4 \pm 0.1$   & - & 0.82   \\       
 I.1 & $13.9 \pm 0.2$    & 0.1& 0.02\\
 II.2 & $44.3 \pm 0.4$ &  0.3  & 0.51\\
II.3 & $76.0 \pm 0.3$   & 0.6 &  1.03\\\hline
\end{tabular}
 \end{center}
\vspace{1mm}
\end{table}

\subsection{Host galaxies}

A meaningful definition of a LGRB host galaxy in our scenarios is not as
straightforward as it might seem at first sight. The na\"ive definition of a
host galaxy, at a given redshift $z$, as being any galaxy $g$ with a LGBR rate
of $r_{\rm GRB}(g,z) > 0$ (or identically, ${\rm SFR}(g,z) > 0$) is not useful
because the resulting host population would include galaxies with arbitrarily
low LGRB rates. Low rates are better understood in statistical terms, as very
low probabilities of producing a LGRB per unit time. This implies that the
corresponding galaxies have a low probability of being observed as host
galaxies. Hence, this na\"ive definition would generate a host galaxies sample
biased to low SFR galaxies. To fix this problem, the definition of a host
galaxy could be based on the number of LGRBs $N(g,z) = r_{\rm GRB}(g,z) \Delta
t$ produced in each galaxy during a time interval $\Delta t$ in its rest frame, so that host
galaxies are only those with $N(g,z) \geq 1$, as in \citet{Cam09}. However this
cut-off, and the resulting population, would be dependent on the more or less
arbitrary choice of $\Delta t$.

Given the above arguments, we preferred instead a probabilistic approach,
defining the likelihood of a galaxy being observed as a LGRB host. This
likelihood is then used as a weight to compute the properties of the observable
host galaxies sample, which would be in this way comparable to the observed
sample. For a galaxy to be detected as a host at least one LGRB must be
detected within it by a high-energy observatory. Then, the galaxy itself must
be detected, usually by optical/NIR telescopes. We would treat the biases
introduced by each of these observations separately.

To model the first bias, we compute the probability of detecting at least one
LGRB within it by any of the high-energy observatories monitoring GRBs. Given that event detection processes follow Poissonian
statistics, if $R_{{\rm GRB},i}(g,z)$ is the contribution of galaxy $g$ at
redshift $z$ to the LGRB rate observed by experiment $i$, the probability of observing at
least one LGRB in a galaxy is

\begin{equation}
\label{phg}
p_{\rm HG}(g,z) = 1 - \exp \left(-\sum_i R_{{\rm GRB},i} \Delta T_i \right),
\end{equation}

\noindent
where $\Delta T_i$ is the time interval of the LGRB search conducted by
experiment $i$. Eqn.~\ref{phg} assumes that the searches by different
observatories are independent. Previous works
estimate the intrinsic LGRB rate in a typical galaxy to be of the order of
$10^{-3}\,{\rm yr}^{-1}$ \citep{Fry99}, hence the LGRB rate observed in the
galaxy by any mission must be lower than this value. Given that current searches for LGRBs
have spanned a few years, $\sum_i R_{{\rm GRB},i} \Delta T_i \ll 1$, and $p_{\rm HG}(g,z)
\simeq \sum_i R_{{\rm GRB},i} \Delta T_i$, which means that the likelihood of a galaxy being
observed as a host is proportional to its contribution to the mean {\em observed}
number of LGRBs,

\begin{equation}
\label{rateobsgal}
p_{\rm HG}(g,z) \simeq  \frac{r_{\rm GRB}(g,z)}{V (1+z)} \frac{dV}{dz} \sum_i p_{{\rm det},i}(z) \Delta T_i \frac{\Delta \Omega_i}{4 \pi},
\end{equation}

\noindent
where $\Delta \Omega_i$ is the sky coverage of the experiment $i$.
Note that, for a fixed $z$, cosmological and observatory dependent factors
in Eqn.~\ref{rateobsgal} become constant, independently of the number of observatories considered, and cancel out in the normalization of the weights. This means that when computing mean HG properties such as mass or SFR as a function of $z$, the
first bias can be modeled simply by weighting the corresponding properties of
each galaxy by its {\em intrinsic} LGRB rate.

In the computation of the integrated properties of the
whole population of host galaxies (like the integrated mass distribution, for
example), also the effects of volume variation, time dilation and detectability
of the different observatories must be taken into account (Pellizza et al. in preparation). In this case it is better to compute first the mass distribution observed by each experiment at each redshift $z$, weighting the galaxies in the corresponding snapshot by $p_{\rm HG}(g,z)$. Second, the integration in $z$ can be performed for each experiment, and the resulting distribution can
be normalized. This has the advantage of avoiding the use of the
values of $\Delta T_i$ and $\Delta \Omega_i$, which are poorly known for
some of the observatories that detected the LGRBs in the sample of \cite{Sav09}.
Only $p_{{\rm det},i}(z)$ for each experiment is needed, which is computed as described in Sect.~\ref{obsrate} for $p_{\rm det,BATSE}(z)$. Finally, these distributions can be combined into a single one by adding them, previously
scaled to the number of LGRBs detected by each observatory. The relevant data
for computing the detectability were taken from \citet{Ste01} for BATSE,
\citet{Gue07} for {\it Swift}, \citet{Fro09} for {\it Beppo-SAX}, \citet{Pel08} for {\it HETE-2} and \citet{Hur92} for {\it Ulysses}. These experiments detected 36 of the 38 LGRBs in the sample of \citet{Sav09}. For statistical reasons, only
experiments that detected at least five LGRBs in the sample were considered. Konus-{\it Wind} and {\it NEAR}-XGRS were discarded because all their LGRBs were
observed also by other experiments above. To be consistent, the observed mass distribution to which the model predictions were compared was constructed from
the data of \citet{Sav09} by adding the individual distributions observed by
each experiment.

The second bias is more difficult to model. Given that the search for host
galaxies is guided by the discovery of the LGRBs themselves and usually done
with a variety of different telescopes and detectors, in different bands of the
electromagnetic spectrum and with different sensitivities, the biases
introduced are unclear. Galaxies with surface brightness below the
detectability threshold of available instruments could introduce a 
bias towards low metallicity galaxies.
An extra problem could be caused by dust obscuration
 which could affect the detection of the  afterglows and produce a bias
towards high metallicity galaxies \citep{Fyn09}.
 Hence, at least these two effects might combine
themselves to determine the detectability of a host galaxy. 
Considering also the fact that we have only  access to the public galaxy
catalogue of the {\it Millennium Simulation} which provides only mean global properties and magnitudes, we adopt the observed integrated stellar mass distribution
 as a tool to apply the combined effect of observational biases to the simulated galaxy sample.
We use the fact that  the observed integrated stellar mass distribution has been affected by observational biases although we cannot
disentangle their individual effects.
Hence, we require the simulated LGRB hosts to reproduce the observed
stellar mass distribution in order to build up the observable simulated 
LGRB hosts. The procedure is explained in detail in next section.
Note that, hereafterin, we will discuss the trends of the observable simulated hosts and the general galaxy population. The former can be compared to the current observed hosts, but, if this later sample
changes due to better or different observational techniques, our simulated
sample should be also consistently modified to match the
new observed stellar mass distribution.

\section[]{LGRB host properties}
\label{hgs}

The largest and most comprehensively uniform sample of host galaxy properties available at
present is that compiled by \citet{Sav09}\footnote{Data available at
http://www.grbhosts.info}. Stellar masses, star formation rates, metallicities,
absolute magnitudes and colours of 46 observed host galaxies up to $z \sim 3$ were
obtained by these authors comparing their spectral energy distributions to
those of synthetic stellar populations, using the method described in
\citet{Gla04}.

A key point of our models is that we require the stellar mass distribution of the predicted host
galaxies to  match that of the observed host galaxies.  The stellar
mass is adopted as the property to be reproduced by the models since it is now
widely accepted that stellar mass is a more fundamental quantity for galaxies
than luminosity \citep[e.g.][]{Kau04}. It could be possible that those LGRB events with no
   detected hosts occured in very low surface brightness galaxies and hence, with low
   stellar masses. As mentioned before the effects of dust could also prevent the
detection of events in high metallicity (and dusty) galaxies. In our models these  host galaxies exist (and are part of the global galaxy population) but the observability cut-off has been determined by using the  current observed stellar mass distribution.
 Fig.~\ref{mass} shows the distribution
of stellar masses of observed host galaxies constructed from the data of \citet{Sav09}, together with those predicted by our scenarios.

As it can be seen from Fig.~\ref{mass}, the scenario which best reproduces the
observed stellar mass distribution is that with $Z_{\rm C}=0.6\, {\rm Z_{\odot}}$
(scenario II.3, see also Table 1). Lower metallicity thresholds predict lower observed stellar
masses for the host galaxies, while no metallicity threshold (scenario I) predicts larger
ones. In the sample of \citet{Sav09}, we find that $85$ per cent of the studied galaxies have
stellar masses over the range $\approx 10^{8.5-10.3} M_\odot$
\citep[see also][]{Cas08}. The probability of getting a host galaxy within this stellar
mass range in scenario II.3 is 88 per cent, while for scenarios I, II.1 and II.2 it is
70, 2, and 43 per cent,  respectively. Then we conclude that scenario II.3 predictions
agree fairly well with observations while others fail, and therefore in the
following sections,  we will focus only on this scenario.

 We stress the fact that
its predictions include the effects of host galaxies observability, hence, a proper
comparison with observations can be made. In order to contribute to the
understanding of the nature of LGRB host galaxies, we will also compare these predictions
with the properties (not weighted by host galaxy observability) of both
the sample of all galaxies with mean cold gas metallicities below $0.6\, {\rm Z_{\odot}}$ (hereafter low
metallicity sample) and the complete galaxy population of the catalogue
of \citet{DeL07}.

\subsection{Stellar mass}

\begin{figure}
\includegraphics[width=0.45\textwidth]{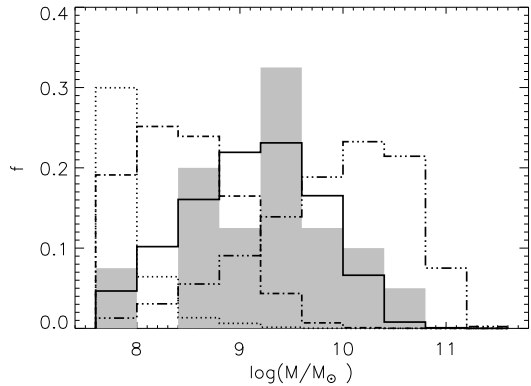}
\caption{Stellar mass distribution of host galaxies in the sample of \citet{Sav09}
(shaded in gray), together with those predicted by scenarios I (dashed-triple dotted line),
II.1 (dotted line), II.2 (dashed-dotted line) and II.3 (solid line).}
\label{mass}
\end{figure}

\begin{figure}
\includegraphics[width=0.45\textwidth]{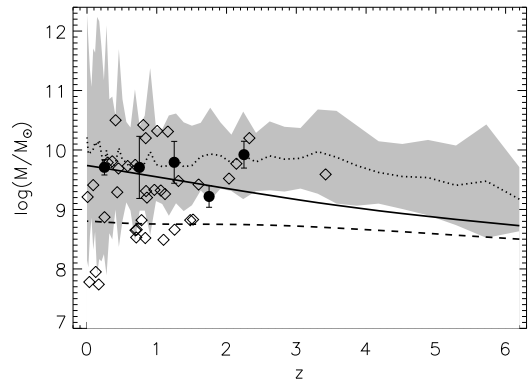}
\caption{Mean  stellar mass of the  host galaxies as a function of redshift predicted by scenario
II.3 (dotted line) with its standard deviation  (shaded grey
band). Open diamonds represent the  observed
host galaxies of \citet{Sav09}, while filled circles correspond to their mean values in redshift intervals
of 0.5. For comparison, we also include the corresponding mean trends for the
low metallicty sample (dashed line) and the complete galaxy population (solid
line).}
\label{massred}
\end{figure}

As a first step towards understanding the nature of host galaxies, we analyse their
stellar masses as a function of redshift. As shown in Fig.~\ref{massred}, the
mean stellar mass of the host galaxies as a function of redshift predicted by scenario II.3
reproduces the observed mean trend quite well. From this
figure, we can also see that host galaxies are, on average, more massive than galaxies in
the complete galaxy population while the latter are more massive than those in
the low metallicity sample. This can be understood taking into account
 that the host galaxy observability is a strong function of the star formation rate,
and that the complete galaxy population in the catalogue of
\citet{DeL07} follows a mass-metallicity relationship \citep[e.g.][]{DeR09}.
Then, the cut-off adopted for the  mean cold gas metallicity to reproduce the observed stellar mass
distribution implies a cut-off in stellar mass since low metallicity galaxies
are, on average, less massive than the general galaxy population (see also Fig. \ref{mass}). However, as
the  observability of a host galaxy depends strongly on its star formation activity and most of
the small galaxies have low star formation rates, the observable host galaxies tend
to be, on average, the  more massive ones among them. As a result, our
observable sample tends to be populated by systems more massive than
those in  the low metallicity sample or in  the complete
galaxy population.

\subsection{Star formation rate}

\begin{figure}
\includegraphics[width=0.45\textwidth]{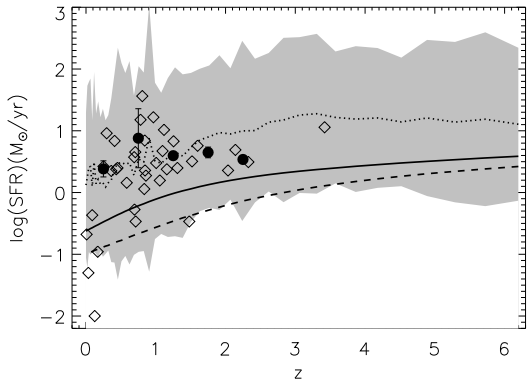}
\caption{Mean  SFR of the host galaxies as a function of redshift predicted by scenario II.3
(dotted line) with its standard deviation  (shaded grey band). Open
  diamonds represent the observed
host galaxies of \citet{Sav09}, while filled circles correspond to their mean values in redshift intervals
of 0.5. For comparison, we also include the corresponding mean trends for the
low metallicty sample (dashed line) and the complete galaxy population (solid
line).}
\label{gsfr}
\end{figure}

In Fig.~\ref{gsfr}, we display the mean SFR of  host galaxies as a function of redshift
predicted by scenario II.3, together with the corresponding mean values for the low metallicity
sample and the complete galaxy population. As it can be seen, the prediction
of scenario II.3 reproduces very well the behaviour of the observed host galaxies. These
have higher SFR than the mean of the complete galaxy population, and much
higher than that of the low metallicity sample. The good agreement between our
scenario II.3 and observations suggests that the observed host galaxies are biased
towards galaxies with stellar masses in the range $10^{9-10} {\rm M_\odot}$, high star
formation activity and relatively low gas metal content, compared to the mean
properties of the complete galaxy population at a given $z$.

\begin{figure}
\includegraphics[width=0.45\textwidth]{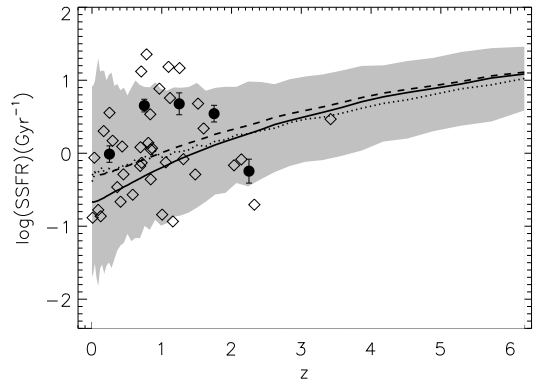}
\caption{Mean  SSFR of the  host galaxies as a function of redshift predicted by scenario II.3
(dotted line) with its standard deviation  (shaded grey band). Open
  diamonds represent the observed
host galaxies of \citet{Sav09}, while filled circles correspond to their mean values in redshift intervals
of 0.5. For comparison, we also include the corresponding mean trends for the
low metallicty sample (dashed line) and the complete galaxy population (solid
line).}
\label{gssfr}
\end{figure}

Regarding the mean specific SFR (SSFR), defined as the ratio between the SFR
and the stellar mass of a given galaxy (Fig.~\ref{gssfr}), host galaxies are predicted to
show SSFRs similar to those in the low metallicity sample, and higher than
those in the complete galaxy population, for $z < 2$. Between $z \sim 2$ and
$z \sim 6$, there are no significant differences in mean  SSFRs among the three
samples, which have larger SSFRs that low redshift galaxies. This can be
understood because, at higher redshift, galaxies have larger gas reservoirs
which can feed stronger star formation activity and are, on average, less
chemically enriched. At low redshift, host galaxies seem to be particularly efficient at
transforming gas into stars. Note, however, that the mean SSFR predicted by
scenario II.3 at low redshift is half an order of magnitude lower than
the mean observed SSFR.

\subsection{Luminosity and colour}

\begin{figure}
\includegraphics[width=0.45\textwidth]{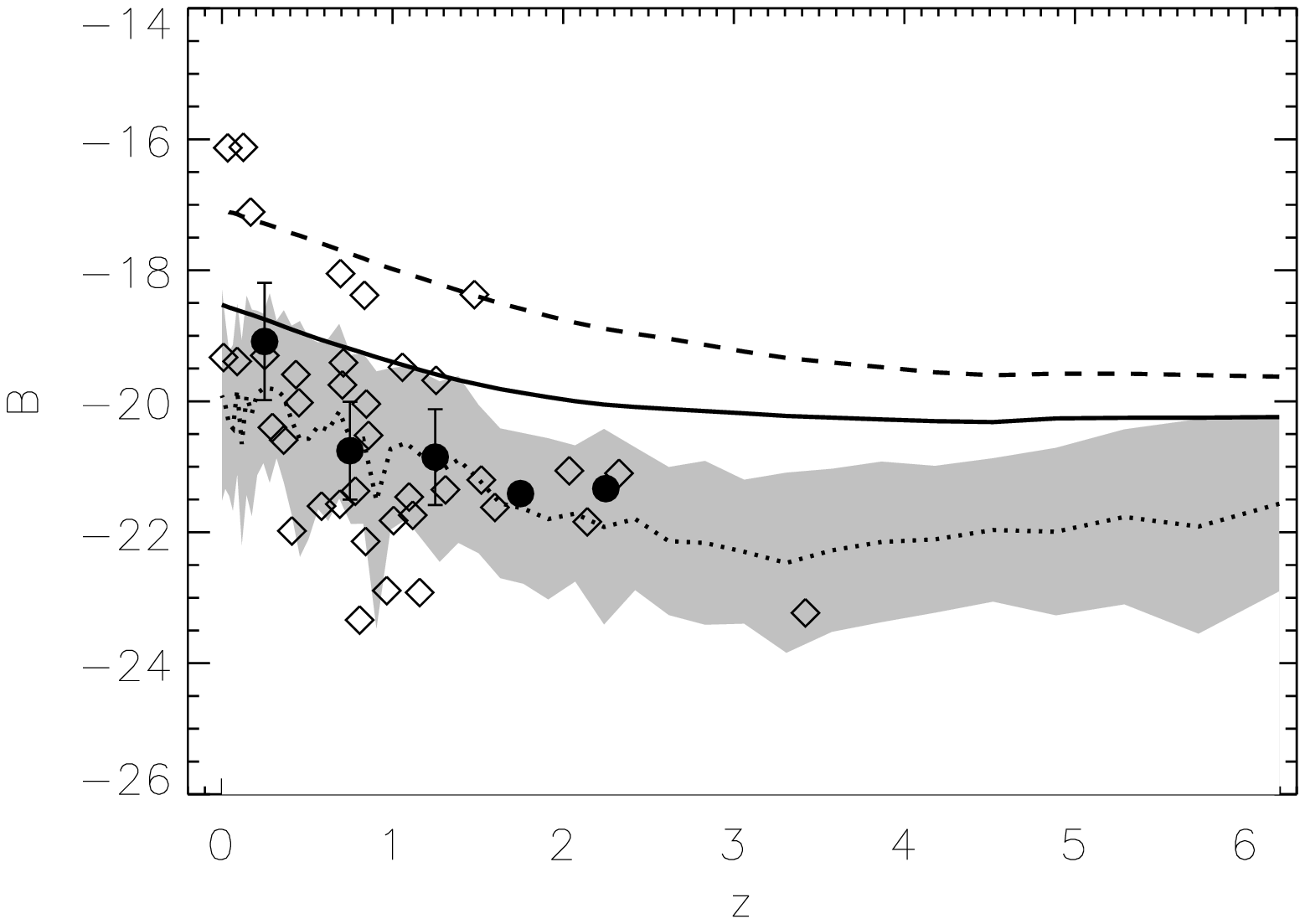}
\includegraphics[width=0.45\textwidth]{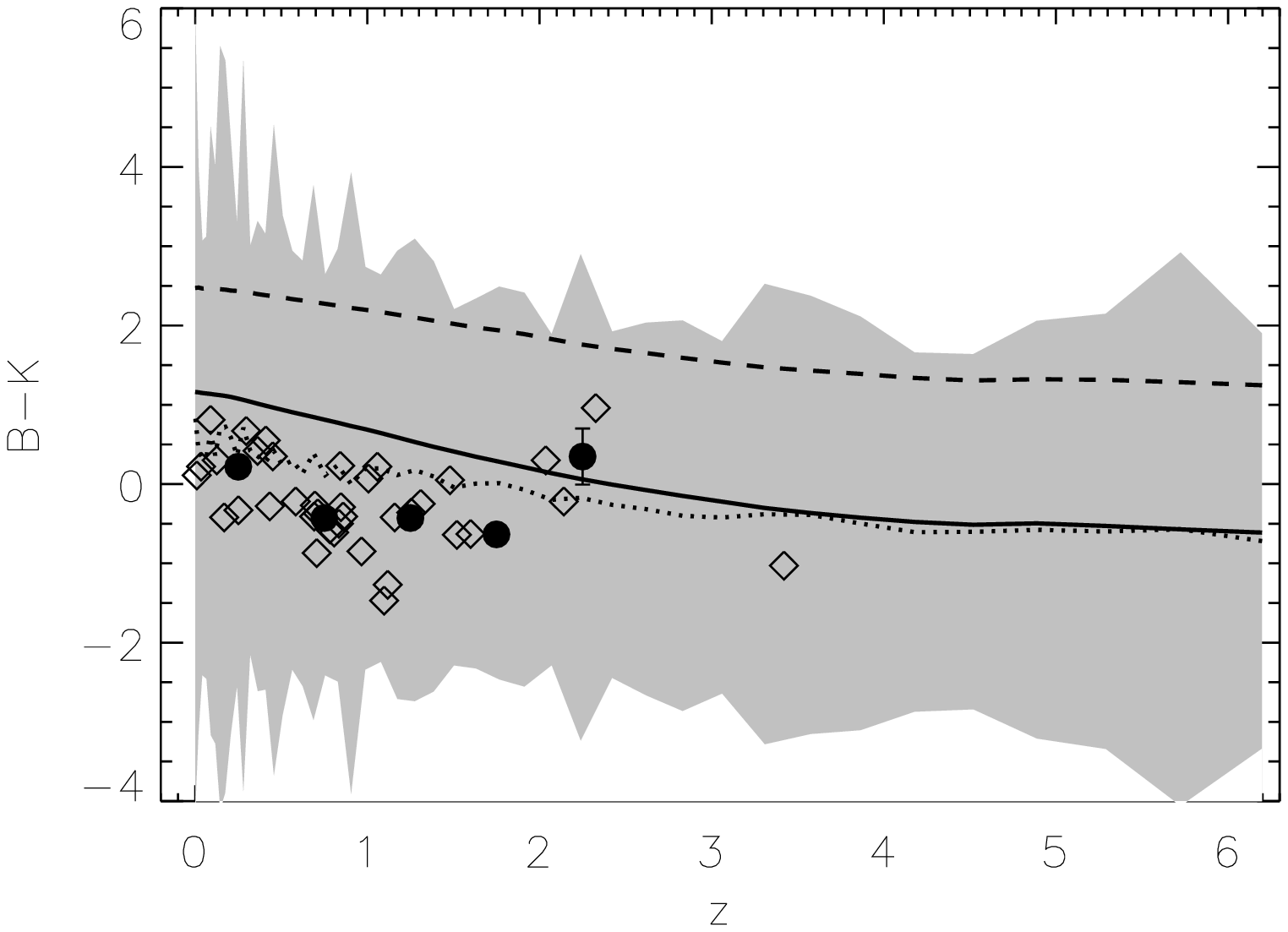}
\caption{Mean $B$-band luminosity (upper panel) and $B-K$ colour (lower panel)
as a function of redshift predicted by scenario II.3 (dotted line) with its
standard deviation 
(shaded grey band). Open diamonds represent the  observed host
galaxies  by \citet{Sav09}, while filled circles
correspond to their mean values in redshift intervals of 0.5. For comparison,
we also include the corresponding mean trends for the low metallicty sample
(dashed line) and the complete galaxy population (solid line).}
\label{colours}
\end{figure}

Observations show that host galaxies tend to be bluer than the general population of
galaxies observed at a given redshift, and fainter than a typical $L^*$ galaxy.
In fact, this trend can be nicely reproduced by our scenario II.3, as shown in
Fig.~\ref{colours}. The  luminosities and colours of the predicted
host galaxies are in excellent
agreement with the observations compiled by \citet{Sav09}. From this figure, we
can see that host galaxies are bluer than the complete galaxy population for $ z < 2$,
but have similar mean colours for higher redshfits. Our scenario predicts that
host galaxies are more luminous systems in the $B$-band, compared to the mean luminosity of
the global galaxy population at all redshifts. 

\subsection{Metallicity}

\begin{figure}
\includegraphics[width=0.45\textwidth]{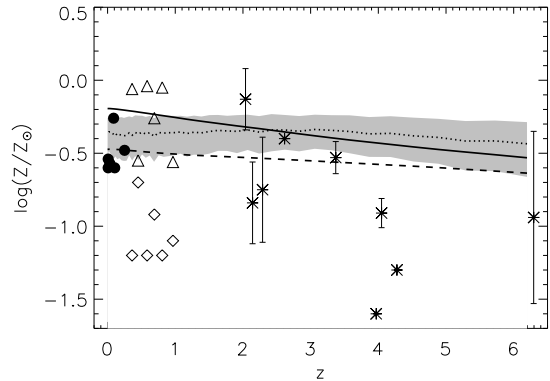}
\caption{Mean cold gas metallicity as a function of redshift predicted by scenario II.3
(dotted line) with its standard deviation (shaded grey band). Filled circles represent observed
metallicities of host galaxies reported by
\citet{Sav09}, measured using the $T_e$ technique. We also include those measured
by the R32
upper branch (triangles) or lower branch (diamonds), and GRB-DLAs (asterisks). For
comparison, we include the corresponding mean trends for the low
metallicty sample (dashed line) and the complete galaxy population (solid
line).}
\label{metal}
\end{figure}

In Fig.~\ref{metal}, we compare the mean cold gas metallicity of the host galaxies
predicted by scenario II.3 to different metallicity estimations reported by
\citet{Sav09}. Since at $z > 0.2$ it is very difficult to distinguish HII
regions with enough resolution, these authors generally measured the optical
luminosity-weighted mean metallicity in a galaxy. For consistency, host galaxies at
$z<0.2$ were treated by them as the rest of the sample (i.e. integrating fluxes
over the whole galaxy). In this figure, we have included metallicities obtained
by using different indicators \citep[see][for more details]{Sav09}. In
some cases, absorption lines in the optical afterglow can be used to obtain the
 metallicity of neutral cold gas along the line of sight of the GRB (the
so-called GRB-DLAs). This is the case of 9 GRB-DLA systems studied by
\citet{Sav06}, all of them at $z > 1.6$. In this case, the metallicity could be
associated more directly to the metallicity of the host galaxy, contrary to QSO-DLAs which
are associated to HI clouds in the intergalactic medium.
 At $z<1$, measurements of metallicity
by \citet{Sav09} are derived from hot gas and, in this case, the lower branch
solution is preferred for the hosts. As shown in Fig.~\ref{metal}, the predicted host galaxies
metallicities are systematically higher than the GRB-DLA metallicities. This
might indicate that LGRBs occur in regions of even lower metallicity than
the mean metallicity of the cold gas of the host galaxies. In this sense, \citet{Nuz07},
adopting similar hypotheses to generate synthetic LGRB populations, but using 
full cosmological simulations where the chemical enrichment of baryons was
consistently followed with redshift, claimed a metallicity threshold of $0.3
Z_{\odot}$ for the progenitor stars in order to reproduce observations. More
detailed information on the metallicity of individual stellar populations,
which also takes into account the inhomogeneities of the interstellar medium, are
needed to improve our understanding of this issue (see \citealt{Pon09} and
Artale et~al. in preparation). Also along these lines, a work by 
\citet{Mod08} compares the metallicity
of GRB hosts associated to Type Ic supernovae (SN Ic) to broad-line SN Ic without GRB
detection at $z < 0.25$. The metallicity of SN-GRB host galaxies are
estimated by computing the nebular oxygen abundance by three different
methods. In this case, they directly probe the environments of the progenitors because either they were measured at
that location or they correspond to metallicities of homogeneous dwarf galaxies. \citet{Mod08}
results are in agreement with the values obtained by \citet{Sav09} and with
our own results at low redshifts.

\section[]{Environment of LGRB hosts}
\label{env}

It is still under discussion if host galaxies are always located in regions of a given
characteristic local density or if, depending on $z$, they  inhabit different
environments. Recently, \citet{Cam09} estimated the cross-correlation function
of their simulated host galaxies finding that these systems tend to map underdense
regions. Observations provide somehow contradictory results. \citet{Bor04}
found host galaxies to inhabit field regions. Lately, \citet{Wai07} suggested that host galaxies
might be biased towards interacting or merging galaxies. It is yet too soon for
observations to provide a robust answer to these questions but theoretical
predictions could positively contribute to this area. The advantage of our
model is that taking into account the observability allows us to make a more
robust prediction.

In this section, we investigate the environment of the simulated host galaxies by means
of three different estimators. The distance to the closest neighbour ($d_1$) of
a host galaxy provides an estimation of the possibility of a host galaxy being interacting with
another galaxy. The distance to the fifth neighbour ($d_5$) of a host galaxy provides an
estimation of the density of the region the galaxy inhabits. Finally, the
central halo virial mass ($M_{\rm vir}$) is an estimator of the global
potential well to which the host galaxy is bound.

In the particular case of the closest neighbour, it is well known that pair
interactions can enhance the star formation activity
\citep[e.g.][]{Lam03,Pat05} and if LGRB progenitors are massive stars this
would imply an increase in the probability of detecting a LGRB event. Since
semi-analytical codes do not model tidally-induced star formation, we modified
the SFR for galaxies with their closest neighbour located at distances smaller
than $0.1 h^{-1}\,{\rm Mpc}$ by a factor of two, motivated by observational and
numerical results \citep[e.g.][]{Lam03,Per06,Per09,DiM08}. However, this
correction does not change any of the trends reported in this work. But the role of
interactions should be further tested with models that properly include their effects.

\begin{figure}
\includegraphics[width=0.45\textwidth]{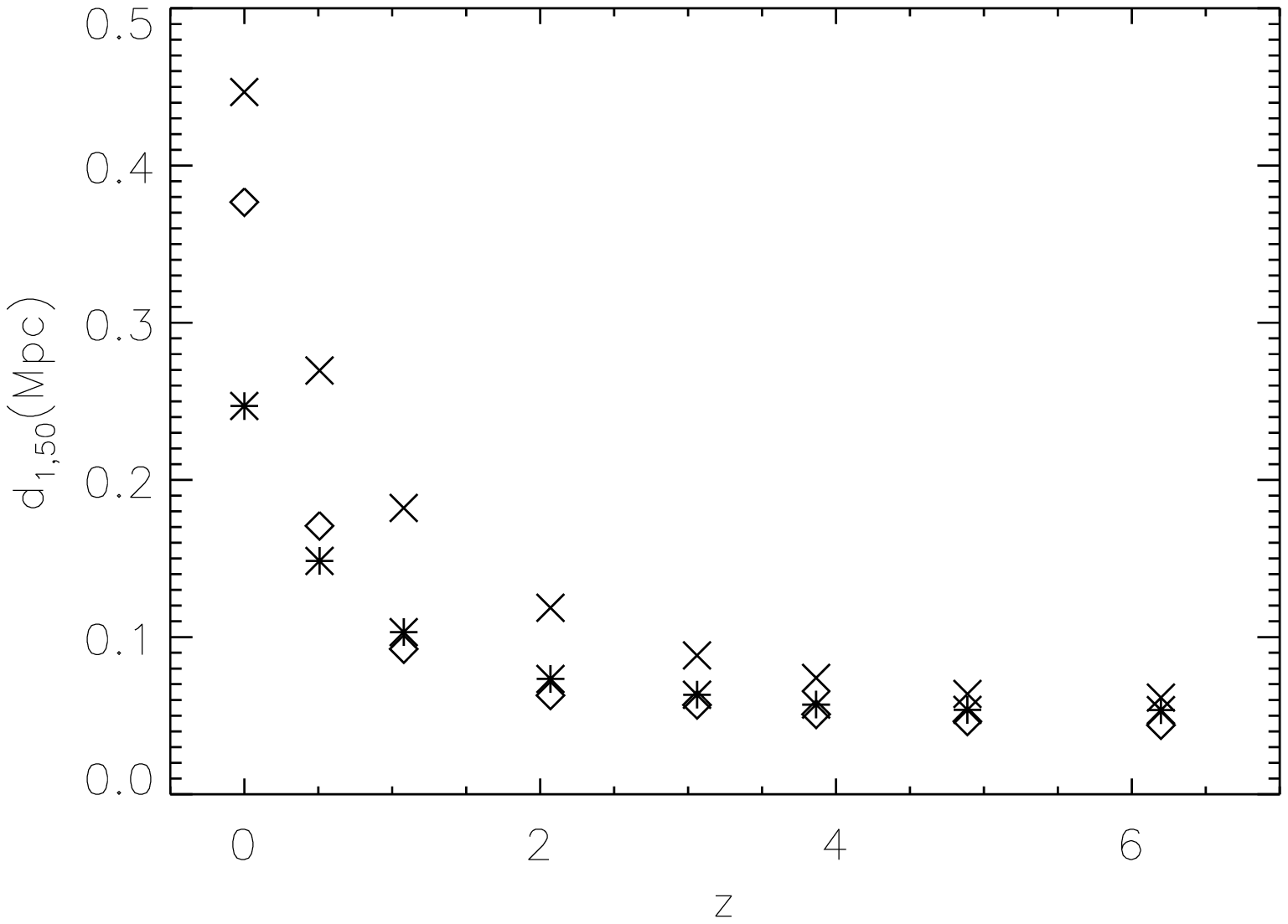}
\includegraphics[width=0.45\textwidth]{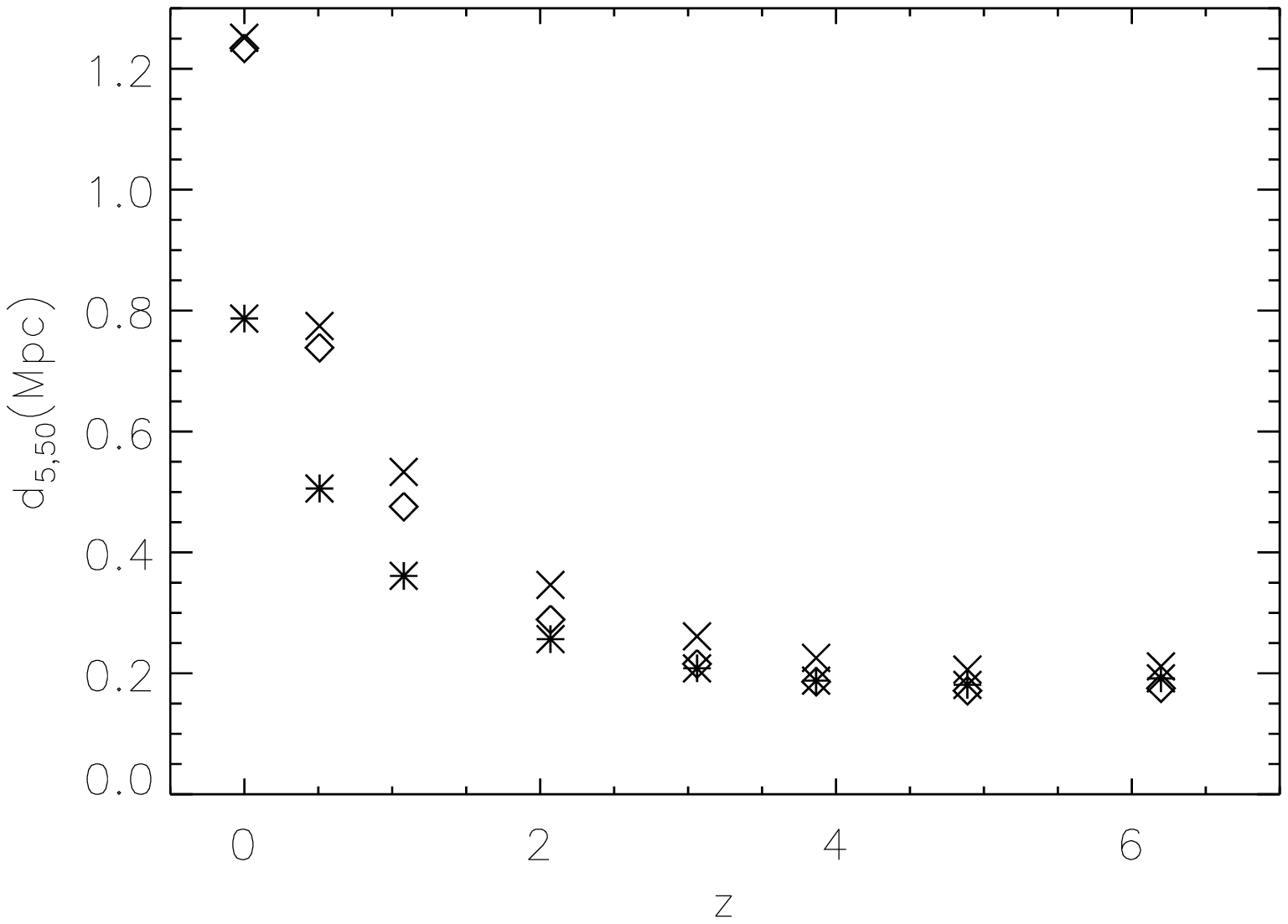}
\includegraphics[width=0.45\textwidth]{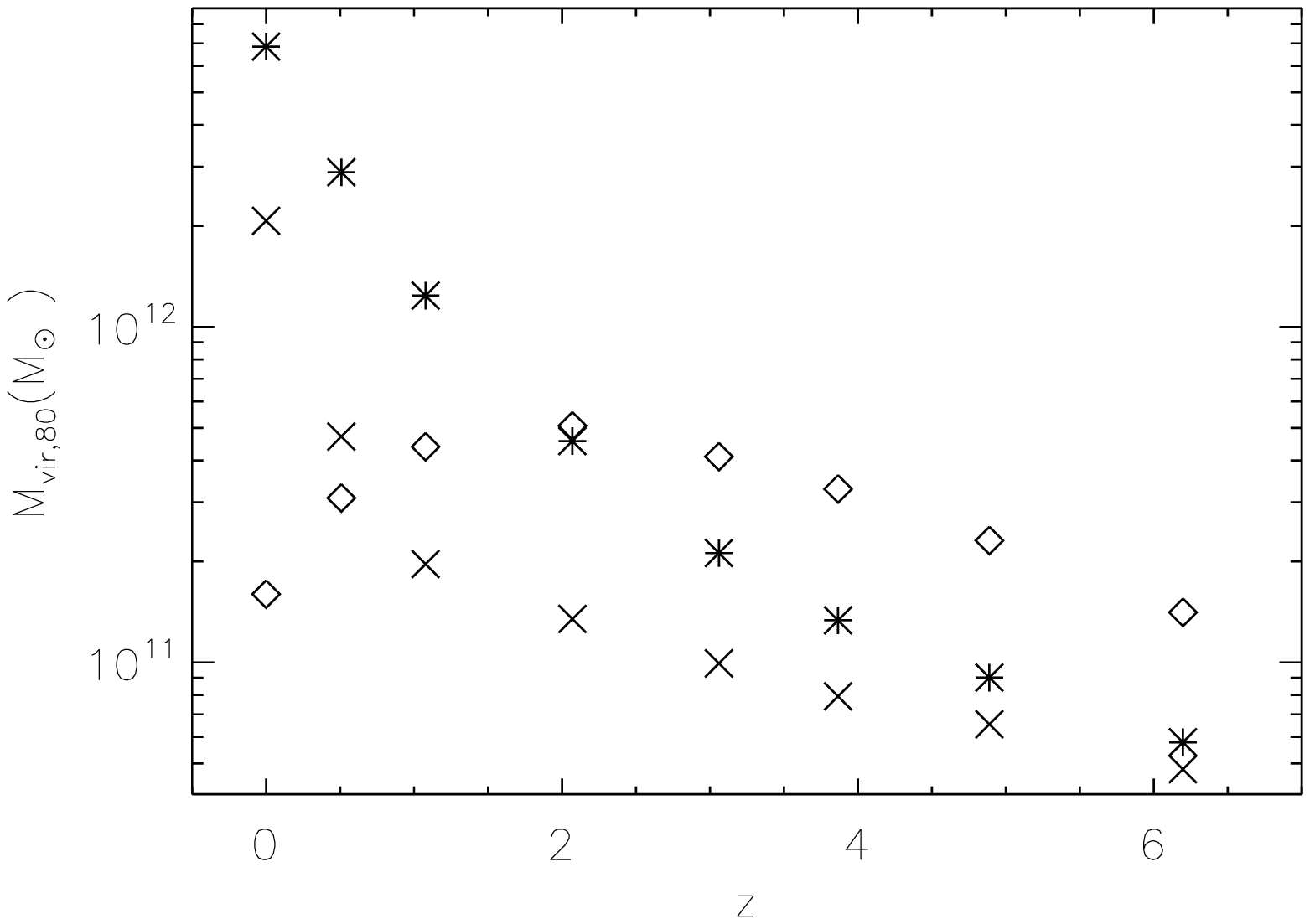}
\caption{Environmental study of host galaxies predicted by scenario II.3 (diamonds), the
complete galaxy population (stars), and the low metallicity galaxy sample
(crosses). {\em Upper panel:} Median distance to the closest neighbour
($d_{1,50}$). {\em Middle panel:} Median distance to the fifth neighbour
($d_{5,50}$). {\em Left panel:} Virial mass of the central halo at the
$80$  percentile ($M_{\rm vir,80}$).}
\label{d1}
\end{figure}

We estimate the cumulative distribution of these three environmental estimators
($d_1,d_5, M_{\rm vir} $) predicted by scenario II.3. To quantify behaviours as
a function of redshift, we determine the values $d_{1,50},d_{5,50}$ and
$M_{\rm vir,80}$ at which the corresponding cumulative fractions reach
$50$ per cent 
in the case of $d_1$ and $d_5$, and $80$ per cent in the case of $ M_{\rm vir}$. In
Fig.~\ref{d1} (upper panel), we show $d_{1,50}$ as a function of redshift
according to scenario II.3, together with the trend for the low-metallicity
sample and the complete galaxy population. At high redshift, the three
samples tend to
have the nearest neighbour at approximately similar distances. We only note a
slight trend for host galaxies to systematically have a closer first neighbour than
galaxies in the other two samples for $z > 1$ . From $z \sim 1$, low
metallicity galaxies tend to have their closest neighbour further away than
either a galaxy in the complete population or a host galaxy. The latter tend to
inhabit lower density regions than galaxies in the complete population. Our
models indicate that observable host galaxies would only have a slightly higher probability to be in
a galaxy pair compared with the general galaxy population at high redshift
(i.e. $50$ per cent  of galaxies have the first neighbour closer than
$\sim 50\,{\rm kpc}$).

The global environment is quantified by $d_{5,50}$. As shown in Fig.~\ref{d1}
(middle panel) at high redshift ($z > 2$), galaxies in the three samples reside
in similar environments. For lower redshift there is a clear trend for galaxies
in the low metallicity sample and host galaxies to inhabit lower density regions than
galaxies in the complete galaxy population. This results agree with the
numerical work of \citet{Cam09}, and also with the observational findings of
\citet{Bor04}, taking into account that the mean redshift of LGRBs observed at
that time was $\langle z \rangle < 2$.

In the case of the virial mass, as shown in Fig.~\ref{d1} (lower panel), we
find that  $80$ per cent  of observable host galaxies have haloes less massive than
$10^{11.0-11.5}\, {\rm M_\odot}$ at any redshift. As the structure forms and groups and
clusters aggregate hierarchically, galaxies tend to inhabit larger dark matter
haloes. However, host galaxies stay within a narrower range of mass haloes. The inversion
in the relation observed in this figure can be understood considering that host galaxies
observability depends strongly on the star formation activity which, in turn,
depends strongly on environment \citep{Pog09}. Hence, from $z \sim 2$ active
star forming galaxies (i.e. those which are important generators of LGRBs)
with mean stellar masses of $\sim 10^{10}\, {\rm M_{\odot}}$  reside again in
slightly smaller dark matter haloes (Fig.~\ref{massred} and Fig.~\ref{gsfr}).
Note that this is not the case for the general low metallicity sample which
although residing, on average, in smaller haloes than the global galaxy
population, tend to systematically inhabit larger ones with decreasing
redshift, as expected in a hierarchical scenario. The observability condition
which is closely linked to the star formation activity produces this kind of
halo downsizing scenario for host galaxies at low redshift, while the mean stellar mass remains
approximately constant.

\section{Conclusions}
\label{con}

Our results indicate that the observed LGRB host galaxies properties can be reproduced
assuming that LGRB progenitors are massive stars  and occur
in galaxies with moderately low mean cold gas metallicities.
The joint requirement for the synthetic LGRBs to reproduce the
observations of BATSE and for the simulated  systems that host them  to reproduce
the  stellar mass distribution of the observed host galaxies
determines the minimum mass for the progenitor star ($m > 75
M_{\odot}$)\footnote{The mass threshold for LGRB progenitors was
computed for the \citet{Cha03} IMF used in the catalogue of \citet{DeL07}, and
would be  54 M$_{\odot}$ if a \citet{Sal55} IMF had been
used, instead.} and the maximum metallicity cut-off ($0.6\,{\rm  Z_{\odot}}$)
for the cold gas metallicity of the simulated host galaxies.
 Our
scenario II.3, which satisfies these conditions, succeeds at reproducing the
dependence on redshift of stellar mass, luminosity, colour, SFR, SSFR
and metallicity of the observed host galaxies compiled by
\citet{Sav09} over the redshift range $0 < z < 3$.

Our main findings are:

\begin{enumerate}

\item
The average stellar mass of host galaxies is higher than the average stellar mass of the
complete galaxy population, and remains within a narrow mass range around
$10^{10}\,{\rm M_\odot}$ , with a very weak trend to lower mass for higher redshifts,  due to the double requirement of low mean gas
metallicity and high star formation rates.
We note that the  dispersion is
approximately 0.5 dex around the $\log$ mean stellar mass.

\item
Compared to the characteristic stellar mass $M^* \sim 10^{10.63}\,{\rm M_{\odot}}$
estimated by \citet{Bel03}, host galaxies are low stellar mass systems (88 per cent of our
host galaxies have stellar masses between $\sim 10^{8.5-10.3}\, {\rm M_\odot}$). However,
compared to the mean stellar mass of the complete galaxy population, host galaxies tend
to be massive galaxies. 

\item
The SFR of host galaxies is larger than that of galaxies in the complete population at
all redshifts. At $z > 2$, the SSFR of observable host galaxies is comparable to the
complete galaxy population, but at lower redshift it is systematically higher.
As a consequence, host galaxies are, on average, bluer than the global galaxy population.
The  dispersion found for  the properties of the observable host galaxies
reflect the different histories of formation of galaxies at a given
stellar mass.
 
\item
Our results support the claims for a metallicity threshold to reproduce
the properties of the current observed host galaxies.
The global metallicity threshold of
$0.6\, {\rm Z_{\odot}}$ we derived from our models is an upper limit for the
metallicity of the LGRB progenitor stars in the case of no dust effects. The comparison of the mean
metallicity of host galaxies predicted by our model with GRB-DLAs observations suggests
that LGRBs might be produced in stars of even lower metallicity as pointed out
by other authors \citep[e.g.][]{Nuz07,Cam09}. At low redshift, our
results are in agreement with SN-GRB local metallicity estimations by
\citet{Mod08}. However, observations
show a large spread in metallicity and there are also observational
uncertainties that can affect its determination. Hence, results should
be taken with caution.

\item
For $1< z < 6$, host galaxies seem to be slightly more likely to be in pairs than
galaxies belonging to the other samples. There is, however, a change of
behaviour for $z < 1$, where the at least 50 per cent of the host galaxies seems to have their
closest neighbour further away than galaxies in the complete population, but
closer than galaxies in the low metallicity sample. We highlight that we are
considering neighbouring galaxies of a mass above $10^9\,M_{\odot}$. This limit 
is set by  the numerical resolution of the {\em Millennium Simulation}. Lower
mass companions could also imprint morphological perturbations and trigger star
formation activity \citep[e.g.][]{Lam03} but we cannot pursue this analysis
further on with this galaxy catalogue.

\item
Regarding global environment, our model suggests that,  at $z > 1$, observable  host
galaxies would preferentially inhabit environments of similar density
to those populated by the general population  and
of slightly higher density than those inhabited by the low metallicity
sample. Towards $z = 0$, observable host galaxies tend to be  progressively located in 
less dense environments, which
becomes as subdense as the regions where low metallicity galaxies reside
\citep[see also][]{Cam09}.

\item 
Our results suggest that  observable host galaxies tend to have
dark matter haloes in the range
$10^{11.0-11.5}\,M_{\odot}$, regardless of redshift, and show a slight signal for halo
downsizing from $z \sim 1$, distinguishing them from the other two samples
which follow the expected halo mass growth in a hierarchical scenario. This
result is consistent with observable host galaxies being systems with  mean
stellar masses approximately constant, regardless of the age of the Universe. 

\end{enumerate}

Our results are mainly a consequence of the joint requirements to have high star formation activity to ensure observability and to reproduce the current distribution of stellar masses of observed host galaxies. 
However, within the current constrains, galaxies with high masses would
be too metal-rich to produce  LGRBs and low mass systems would have low probability of
being observed. If the observed  sample were modified by
the incorporation of other galaxies, for example dusty hosts, then our model would need 
to be readjusted accordingly.

\section*{Acknowledgments}
We thank M.E. De Rossi and Gerard Lemson for helping us to manage the {\em
Millennium Simulation}, and Sandra Savaglio for her useful comments and suggestions. 
This work was partially supported by grants PICT
2005-32342, PICT 2006-245 Max Planck, and PICT 2006-2015 from Argentine ANPCyT.

\bsp

\label{lastpage}

\end{document}